\documentclass[12pt]{article}
\usepackage{graphics}
\usepackage{epsfig}
\usepackage{amsmath,amssymb}
\usepackage{multirow}

\begin{document}

\setlength{\textheight}{240mm}
\voffset=-15mm
\baselineskip=20pt plus 2pt
\renewcommand{\arraystretch}{1.6}

\begin{center}

{\large \bf On the energy density of linearly polarized, plane gravitational wave}\\
\vspace{5mm}
\vspace{5mm}
I-Ching Yang  \footnote{E-mail:icyang@nttu.edu.tw}

Department of Applied Science, National Taitung University, \\
Taitung 95002, Taiwan (R.O.C.)\\

\end{center}
\vspace{5mm}

\begin{center}
{\bf ABSTRACT}
\end{center}
In this article, the energy density of plane gravitational wave is studied by using Einstein and M{\o}ller's
prescription of energy-momentum pseudotensors. The linearly polarized plan gravitational wave solution 
of Einstein field equation, which has been defined by Bondi et al., is represented by four kinds of different 
coodrinates. The energy distribution of gravitational wave solution in Einstein and M{\o}ller's prescription  
are obtained. Particularly the energy component is zero in null coordinates.

\vspace{2cm}
\noindent
{PACS Nos.: 04.20.Cv, 04.30.Nk \\}

\vspace{5mm}
\noindent

\newpage

Since Einstein~\cite{E16} has predicted the existence of gravitational wave in 1916, a century after 
the first direct observation of gravitational waveshad been made by LIGO and Virgo scientific 
collaboration~\cite{A16}.  But actually, in 1905 Poincar\'{e}~\cite{P05} first proposed the existence 
of gravitational wave and suggested gravitational waves would be generated by accelerating masses 
in the same way electromagnetic waves are generated by accelerating charges.
To consider the weak-field approximation
\begin{equation}
g_{\mu \nu} = \eta_{\mu \nu} + h_{\mu \nu}  ,
\end{equation}
where $\eta_{\mu \nu}$ is Minkowski metric and $\left| h_{\mu \nu} \right| \ll 1$,
Einstein~\cite{E16} obtained the gravitational wave equation in vaccum from Einstein field equations
\begin{equation}
\square^2  h_{\mu \nu} =0 
\end{equation}
with the harmonic gauge condition $\partial^{\mu} h_{\mu \nu} =0$.
It resemble to the electromagnetic wave equation in vaccum
\begin{equation}
\square^2 A_{\mu} = 0
\end{equation}
with the Lorentz gauge condition $\partial^{\mu} A_{\mu} =0$.
Because of a great many similarities between gravitation and electromagnetism, 
it should be therefore expected that Einstein field equations, like  Maxwell field equation, have 
radiative solutions. 

In theory of electromagnetism, the electromagnetic waves can transport energy, so it is believed
that the gravitational waves also can transport energy in general relativity. 
However, the notion of energy is one of the oldest and most controversial problems in general relativity.
Nester et al.~\cite{CNC} show that quasilocal energy-momentum can be
obtained from the Hamiltonian.
\begin{equation}
H({\bf N}, \Sigma)  =  \oint_{\partial \Sigma} {\cal B}({\bf N}) 
= - \frac{1}{2\kappa} \oint_{\partial \Sigma}  N^{\mu} {\cal U}^{\nu \lambda}_{\mu} dS_{\nu \lambda}  .
\end{equation}
Hence, energy-momentum pseudotensors, which are associated with a legitimate Hamiltonian boundary 
term,  are acceptable. So, in this article, Einstein and M{\o}ller's prescriptions of energy-momentum 
pseudotensors are used to study the energy density of plan gravitational waves, which are the simplest 
solutions of wave equation.   

As a compromise between realism and complexity, the linearly polarized, plane gravitational wave 
solution, which has been defined by Bondi et al.~\cite{B59, MTW73}, have been chosen in Cartestian
coordinates $(t,x,y,z)$
\begin{equation}
ds_{\rm (I)}^2 = dt^2 - L^2 \left( e^{2\beta} dx^2 + e^{-2\beta} dy^2 \right) - dz^2  .
\end{equation}
In order to study wave propagation, the light-cone coordinates are always used to express the mathematical
formula of traveling wave. In the light-cone coordinates $(u,v,x,y)$, the line element will become~\cite{MTW73}
\begin{equation}
ds_{\rm (II)}^2 = du dv - L^2 \left( e^{2\beta} dx^2 + e^{-2\beta} dy^2 \right)   .
\end{equation}
Here  
\begin{eqnarray}
u & = & (t+z)/ \sqrt{2}   ,  \\
v & = & (t-z)/ \sqrt{2}   ,
\end{eqnarray}
and $L$ and $\beta$ are functions of  $u$.  
In addition, the so-called plane-fronted gravitational wave with parallel rays (pp-wave) was introduced 
by  Ehlers and Kundt~\cite{EK62} in Brinkmann coordinates $(U,V,X,Y)$ as 
\begin{equation}
ds_{\rm (III)}^2 = 2 H(U,X,Y) dU^2 + 2 dU dV - dX^2 - dY^2  .
\end{equation}
By suitable coordinate transformations
\begin{eqnarray}
X & = & x L e^{\beta}  , \\
Y & = & y L e^{-\beta}  , \\
U & = & \sqrt{2} u  ,  \\
V & = & \sqrt {2} v - x ( L e^{\beta})' - y ( L e^{-\beta})' ,
\end{eqnarray}
the Eq.(6) can be rewritten in the form which is similar to Eq.(9)
\begin{equation}
ds_{\rm (III)}^2 = 2(Y^2 - X^2) \frac{F(U)}{2} dU^2 + 2 dU dV - dX^2 - dY^2   ,
\end{equation}
where 
\begin{equation}
H(U,X,Y) = \frac{Y^2 - X^2}{2} F(U)  .
\end{equation}
And lastly, Bonner~\cite{B69} has described an exact solution of an infinitely straight beam of light 
in the following form
\begin{equation}
ds_{\rm (IV)}^2 = dT^2 - dX^2 - dY^2 - dZ^2 + m (dT - dZ)^2  ,
\end{equation}
where  $m(T,X,Y,Z) = H(U,X,Y)$. So the null coordinate $(U, V)$ is transfered to new coordinates 
$(T, Z)$, named Brinkmann-Cartesian coordinates, by 
\begin{eqnarray}
T & = & (V+U)/ \sqrt{2}  ,  \\
Z & = & (V-U)/ \sqrt{2}  .
\end{eqnarray}
Here, the line element of the gravitational wave solution is represented by four kinds of different 
coordinates.

In physics, energy-momentum, associated with a continuity equation in the differential form 
$\partial_{\mu} T^{\mu \nu} = 0$,  is regarded as the most fundamental conserved quantity.  
However, attempts at identifying an energy-momentum density for gravity, led only to the  
energy-momentum complex which is {\it pseudotensor}
\begin{equation}
\Theta^{\mu}_{\nu} = \sqrt{-g} \left( T^{\mu}_{\nu} +t^{\mu}_{\nu} \right)  
\end{equation}
as the total energy-momentum of matter $ T^{\mu}_{\nu}$ and gravitational field $t^{\mu}_{\nu}$.
After the expression proposed by Einstein~\cite{E15,T} for the energy-momentum complexes with the 
gravitational field 
\begin{equation}
\frac{1}{\sqrt{-g}} \frac{\partial}{\partial x^{\mu}} \left( \sqrt{-g}t^{\mu}_{\nu} \right)
\equiv - \frac{1}{2} \frac{\partial g_{\mu \alpha}}{\partial x^{\nu}} T^{\mu \alpha} ,
\end{equation}
various physicists, such as Landau and Lifshitz~\cite{LL}, Papapetrou~\cite{P48}, Bergmann~\cite{BT53}, 
Weinberg~\cite{W72} and M{\o}ller~\cite{M58}, had given different definitions for the energy-momentum 
complexes.  Particularly, the energy-momentum complex will satisfies the continuity equation
\begin{equation}
\frac{\partial \Theta^{\mu}_{\nu}}{\partial x^{\mu}} = 0
\end{equation}
Mathetamtically, antisymmetric $ {\cal U}^{\mu \rho}_{\nu} $ in their two indices $\mu$ and $\rho$
would be introduced by  
\begin{equation}
\Theta^{\mu}_{\nu} \equiv \frac{\partial {\cal U}^{\mu \rho}_{\nu}}{\partial x^{\rho}}  
\end{equation}
So, the definition of the Einstein energy-momentum complex~\cite{E15,T} is exhibited as
\begin{equation}
{}_{\rm E}\Theta^{\mu}_{\nu} = \frac{1}{16\pi} \frac{\partial H^{\mu \sigma}_{\nu}}{\partial x^{\sigma}}  ,
\end{equation}
where the Freud's superpotential is 
\begin{equation}
H^{\mu \sigma}_{\nu} = \frac{g_{\nu \rho}}{\sqrt{-g}} \frac{\partial}{\partial x^{\alpha}}
\left[ \left( -g \right) \left( g^{\mu \rho} g^{\sigma \alpha} - g^{\sigma \rho} g^{\mu \alpha} \right) \right]  ,
\end{equation}
and the definition of the M{\o}ller energy-momentum complex~\cite{M58}
\begin{equation}
{}_{\rm M}\Theta^{\mu}_{\nu} = \frac{1}{8\pi} \frac{\partial \chi^{\mu \sigma}_{\nu}}{\partial x^{\sigma}}  ,
\end{equation}
where the M{\o}ller's superpotential is 
\begin{equation}
\chi^{\mu \sigma}_{\nu} = \sqrt{-g} \left( \frac{\partial g_{\nu \alpha}}{\partial x^{\beta}} 
- \frac{\partial g_{\nu \beta}}{\partial x^{\alpha}} \right)  g^{\mu \beta} g^{\sigma \alpha}   .
\end{equation}
Here these gravitational wave solutions, which are represented as Eq.(5), (6), (9) and (16) in four differential
coordinates, are considered to investigate their energy distribution.  According to the definition of energy-momentum
complexes of Einstein and M{\o}ller, all components of the Freud's and M{\o}ller's superpotential ${\cal U}^{0i}_0$
is shown in Table 1.
\begin{table}[h]
\begin{center}
\caption{All components of superpotential ${\cal U}^{0i}_0$ }
\begin{tabular}{|c|l|l|} 
\hline cooedinates & Einstein $H^{0i}_0$ & M{\o}ller $\chi^{0i}_0$ \\ \hline
\multirow{3}{*}{ $(t,x,y,z)$} & $H_0^{01} = 0  $ & $\chi_0^{01} = 0$  \\
 & $H_0^{02} = 0$ & $\chi_0^{02} = 0 $  \\
 & $H_0^{03} = -4L \frac{\partial L}{\partial z}$ & $\chi_0^{03} = 0$ \\ \hline
\multirow{3}{*}{$(u,v,x,y)$}  & $H^{01}_0 = L \frac{\partial L}{\partial u}$ & $\chi_0^{01} = 0$  \\  
 & $H_0^{02} = 0$ & $\chi_0^{02} = 0 $  \\
 & $H_0^{03} = 0$ & $\chi_0^{03} = 0$ \\ \hline
\multirow{3}{*}{$(U,V,X,Y)$}  & $H_0^{01} = 0  $ & $\chi_0^{01} = 0$  \\  
 & $H_0^{02} = 0$ & $\chi_0^{02} = 0 $  \\
 & $H_0^{03} = 0$ & $\chi_0^{03} = 0$ \\ \hline
\multirow{3}{*}{$(T,X,Y,Z)$}  & $H^{01}_0 = \frac{\partial m}{\partial X}$ & $\chi_0^{01} = \frac{\partial m}{\partial X}$ \\   
 & $H_0^{02} = \frac{\partial m}{\partial Y}$ & $\chi_0^{02} = \frac{\partial m}{\partial Y}$  \\
 & $H_0^{03} = m \frac{\partial m}{\partial Z}$ & $\chi_0^{03} = \frac{\partial m}{\partial T}+ \frac{\partial m}{\partial Z}$ \\ \hline
\end{tabular}
\end{center}
\end{table}

Finally, the energy distribution of gravitational wave solutions in Einstein and M{\o}ller's prescriptions are obtained, 
and the energy component $\Theta^0_0$ of Einstein and M{\o}ller energy-momentum complexes are shown in Table 2.
The energy component ${}_{\rm E}\Theta^0_0$ and  ${}_{\rm M}\Theta_0^0$ are both equal to zero in null coordinates 
$(u,v,x,y)$ and $(U,V,X,Y)$, but not in non-null coordinates $(t,x,y,z)$ and $(T,X,Y,Z)$. However, in my eaeier 
article~\cite{Y12}, the energy components of Einstein energy-momentum complex for the static spherically symmetric 
space-time with the generalized PG Cartesian coordinates, Kerr-Schild Cartesian coordinates, and Schwarzschild Cartesian 
coordinates are the same. General relativity was introduced along with the principle that not only is no coordinate system 
preferred, but that any arbitrary coordinate system would do.  
\begin{table}[h]
\begin{center}
\caption{The energy component  $\Theta^0_0$ of energy-momentum complexes}
\begin{tabular}{|c|l|l|} 
\hline coordinates & ${}_{\rm E}\Theta^0_0$ & ${}_{\rm M}\Theta_0^0$ \\ \hline
$(t,x,y,z)$ & $ -\frac{1}{4\pi} \left[  L \frac{\partial^2 L}{\partial z^2} + \left( \frac{\partial L}{\partial z} \right)^2 \right]  $
& $ 0$  \\  \hline
$(u,v,x,y)$  & $0$ & $0$  \\  \hline
$(U,V,X,Y)$  & $0 $ & $0$  \\   \hline
$(T,X,Y,Z)$  & 
$-\frac{1}{16\pi} \left[  \nabla_{XY}^2 m + m  \frac{\partial^2 m}{\partial Z^2} + 
\left( \frac{\partial m}{\partial Z} \right)^2  \right]  $  &
$-\frac{1}{8\pi} \left[  \nabla_{XY}^2 m +  \frac{\partial^2 m}{\partial Z^2}  +
\frac{\partial^2 m}{\partial Z \partial T}  \right]  $  \\    \hline
\end{tabular}
\end{center}
\end{table}
But, the energy component  $\Theta^0_0 =0$ in null coordinates shows this choice of coordinates is particular. It means 
that the energy of gravitational wave can't be detect in null coordinates because the observer is moving with the gravitational 
wave. Therefore, the choice of null coordinates should be unique on the exploration of gravitational wave.

\end{document}